\documentclass[preprints,article,accept,moreauthors,pdftex]{mdpi}

\usepackage{upgreek}
\firstpage{1}
\pubvolume{0}
\issuenum{1}
\articlenumber{5}
\pubyear{2019}
\copyrightyear{2019}
\history{Received: 29 December 2018; Accepted: 5 February 2019; Published: date}
\newcommand{\fermi}{{Fermi}}
\newcommand{\gr}{$\gamma$-ray}

\newcommand{\ssr}{Space Sci. Rev.}

\Title{Searching for Gamma-Ray Millisecond Pulsars: Selection of Candidates Revisited}

\Author{Xuejie Dai $^{1,2,*}$, Zhongxiang Wang $^{1}$ and Jithesh Vadakkumthani $^{3}$}

\address{%
$^{1}$ \quad Shanghai Astronomical Observatory, Chinese Academy of Sciences, Shanghai 200030, China; wangzx@shao.ac.cn\\
$^{2}$ \quad Graduate University of the Chinese Academy of Sciences, No. 19A, Yuquan Road, Beijing 100049, China\\
$^{3}$ \quad Inter-University Centre for Astronomy and Astrophysics, Pune 411 007, India; jitheshthejus@gmail.com}

\corres{Correspondence:daixj@shao.ac.cn}
\abstract{We are starting a project to find $\gamma$-ray millisecond
pulsars (MSPs) among the unidentified sources detected by the Large Area
Telescope (LAT) onboard the {Fermi Gamma-Ray Space Telescope (Fermi)},
by radio observations. The selection of good candidates from analysis of the
LAT data is an important part of the project. Given that there is more than 10 years worth
of LAT data and the advent of the newly released LAT 8-year point source list (FL8Y), we have
conducted a selection analysis, on the basis of our previous analysis,
and report the results here. Setting the requirements for the unidentified
sources in FL8Y of Galactic latitudes $|b|>$ 5$^\circ$ and curvature
significances
$>$3$\sigma$, there are 202 sources with detection signficances
$>$6$\sigma$. We select 57 relatively bright ones (detection
significances $>$15$\sigma$) and analyze their 10.2 years of LAT
data. Their variability is checked to exclude variable sources
(likely~blazars), test statistic maps are constructed to avoid
contaminated sources, and
curvature significances are re-obtained and compared to their $\gamma$-ray
spectra to exclude non-significant sources. In the end, 48 candidates
are found. Based on the available information, mostly from multi-wavelength
studies, we discuss the possible nature of several of the candidates. Most of
these candidates
are currently being observed with the 65-meter Shanghai Tian Ma Radio Telescope.}

\keyword{Pulsars; gamma-rays; multi-wavelength}

\begin{document}
%%%%%%%%%%%%%%%%%%%%%%%%%%%%%%%%%%%%%%%%%%

%%%%%%%%%%%%%%%%%%%%%%%%%%%%%%%%%%%%%%%%%%
%\setcounter{section}{-1} %% Remove this when starting to work on the template.
%\section{How to Use this Template}
%The template details the sections that can be used in a manuscript. Note that the order and names of article sections may differ from the requirements of the journal (e.g., the positioning of the Materials and Methods section). Please check the instructions for authors page of the journal to verify the correct order and names. For any questions, please contact the editorial office of the journal or support@mdpi.com. For LaTeX related questions please contact latex@mdpi.com.
%The order of the section titles is: Introduction, Materials and Methods, Results, Discussion, Conclusions for these journals: aerospace,algorithms,antibodies,antioxidants,atmosphere,axioms,biomedicines,carbon,crystals,designs,diagnostics,environments,fermentation,fluids,forests,fractalfract,informatics,information,inventions,jfmk,jrfm,lubricants,neonatalscreening,neuroglia,particles,pharmaceutics,polymers,processes,technologies,viruses,vision

\section{Introduction}

The Large Area Telescope (LAT), onboard the {Fermi Gamma-Ray Space
Telescope (Fermi)}, has been monitoring the whole sky for more than 10 years.
Because of its great capabilities, we have entered
a great era where thousands of \gr\ sources have been detected and identified,
allowing us to study different types of objects and their high-energy emission
processes in detail. From these studies, it~has been found that most of
detected sources belong to the blazar class of Active Galactic Nuclei
(AGN;~\cite{3fagn15}). In the Milky Way,  pulsars are the dominant
\gr\ sources \cite{2fpsr13,3fgl15}. Thus far, 234 pulsars have been
identified to
have \gr\ emission by LAT $\footnote{\url{https://confluence.slac.stanford.edu/display/GLAMCOG/Public+List+of+LAT-Detected+Gamma-Ray+Pulsars}}$, and nearly half
of them are millisecond pulsars~(MSPs).

An effective way to identify \gr\ pulsars is to search for pulsed emission
at radio frequencies in LAT sources. The Shanghai Tian Ma Radio Telescope, a newly built,
65 meter diameter telescope with observing frequencies of 1.25--50 GHz, is
capable of carrying out such searches. For the purpose of finding new
\gr\ pulsars, previously
we had conducted selections of candidate \gr\ MSPs among \fermi\ LAT
unidentified sources \cite{dai+16,dai+17}. The selections were based on
the \fermi\ LAT third source catalog (3FGL; \cite{3fgl15}), which contains
nearly 1000 unidentified sources among 3033 sources, obtained from
the first four years of all-sky monitoring. Our main selection criteria
are (1) Galactic latitudes $|b|>5^{\circ}$; (2) non-variable; (3)
curvature significance {(Signif\_curve in 3FGL; \cite{3fgl15})}
of a source's emission greater than 3$\sigma$
{(see also Section \ref{subsec:sa}).}
{The first is to avoid the relatively crowded Galactic plane.
We note that, among 131 young (i.e., non-MSP) $\gamma$-ray pulsars, only
19 have magnitudes of Galactic latitudes greater than 5$^{\circ}$;
therefore, there is only small chance that we also selected candidate
young pulsars.  The latter two help to distinguish a candidate from
blazars. Pulsars have stable emission, generally described with
a power law with an exponential cutoff (i.e., some degree of curvature
in the spectra), in contrast to the highly variable and power-law like emission
from blazars (e.g., see \cite{wil+14,3fagn15}).}
Using these criteria, we selected 101 sources from 3FGL and found 52 candidate
MSPs; 49 were rejected, due to low detection significances
(average\_sig $<6\sigma$; \cite{3fgl15}), confusion or mixed with extended
bright emission regions, based on
the Test Statistic (TS) maps we calculated \cite{dai+16}, or low curvature
significances based on spectra we obtained \cite{dai+16,dai+17}.
Among the candidates, two sources,
3FGL J0514.6$-$4406 and J1946.4$-$5403, have already been identified as the
pulsars PSR J0514$-$4407 \cite{bha2017} and PSR J1946$-$5403 \cite{ray+16}.

Now, more than 10 years of LAT data have been collected and
the preliminary LAT 8-year point source list (FL8Y)\footnote{\url{https://fermi.gsfc.nasa.gov/ssc/data/access/lat/fl8y/}}
has been released, in which more than 5000 sources are listed, with their
source locations and spectral properties provided. Given these sources,
re-analysis for the source selection is warranted, which will confirm
our previous selection of candidates, add new candidates, and also help to
check the improvement of the new data catalog. We, thus, conducted the analysis
based on the FL8Y and report the results here.

\subsection*{FL8Y Target Selection}

\begin{figure}[H]
\centering
\includegraphics[width=9 cm]{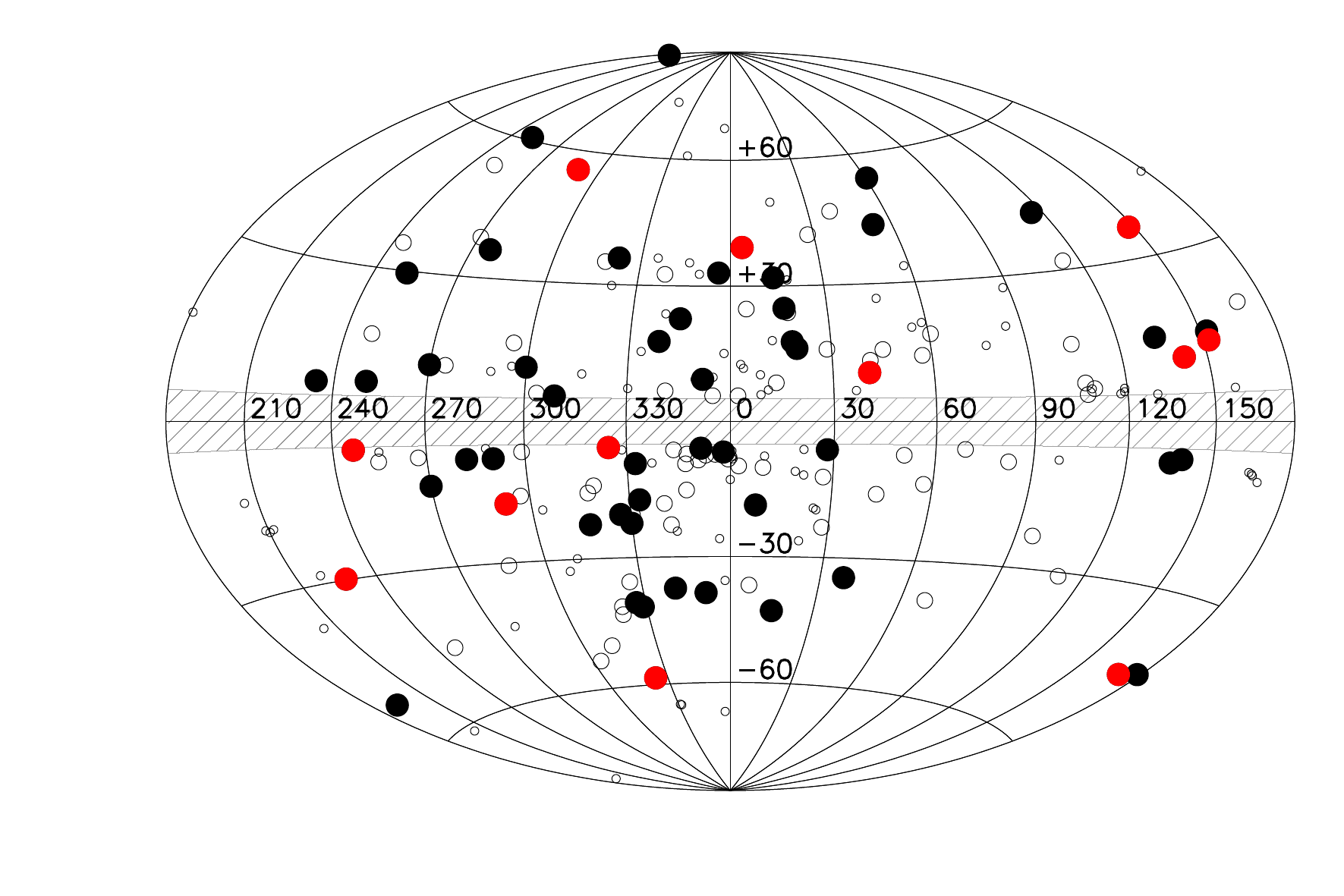}
\caption{Galactic positions of the 202 sources selected from
the preliminary LAT 8-year point source list (FL8Y) catalog. Three~sizes of circles are used to indicate
detection significances of >15$\sigma$, 10--15$\sigma$, and 6--10$\sigma$,
respectively. We~analyzed the data for the first set of 57 sources (marked
by filled circles), and 12 of them were found to be new candidates (red filled
circles).} %%%\textcolor{red}{(there is no red circles in the figure.)}}
\label{fig:sky}
\end{figure}

We selected targets from the unassociated sources in FL8Y using the two
criteria: Having Galactic latitudes of $|b|>$ 5$^\circ$ and having curved
spectra with PLEC\_SigCurv $>$ 3$\sigma$.
No variability information is provided in FL8Y, and we had to obtain it from
our own analysis (see below, in Section \ref{sec2.2}). There are 202 sources with
detection significance Signif\_avg $ >$ 6$\sigma$. Their positions, in
Galactic coordinates, are shown in Figure~\ref{fig:sky}.
Different sizes of circles are used to indicate the detection significances
(the~larger, the higher). In this analysis, we selected those
whose Signif\_avg $>$ 15$\sigma$ (marked with filled circles in
Figure~\ref{fig:sky}). There are 57 sources, with 37 of them
 included in our previous analysis,
and were found to be good candidates \cite{dai+16,dai+17}.
We, therefore, mainly focused on analysis of the data for the \mbox{20 new sources.}

\section{\fermi\ LAT Data and Data Analysis}
\vspace{-6pt}

\subsection{LAT Data}

LAT is a pair-production telescope with a wide field-of-view,
recording the direction, energy, and~arrival time of each
$\gamma$-ray photon in the energy range of from below 20 MeV to
more than \mbox{300 GeV \cite{atw+09}.} It scans the whole sky every three hours.
In this analysis, we used approximately 10.2~yrs of LAT data selected
from the Pass 8 database. The time period is from 4 August 2008  15:43:39 to
\mbox{31 October  2018} 23:56:35 (UTC). For each target, we extracted data in a region
centered at the target's position with a size of 15$^\circ$ {in radius}.
Because of the relatively large uncertainties of the instrument response
function of the LAT in the low energy range (below 200 MeV),
we chose the data in the energy range from 200 MeV to 300 GeV. Events
with zenith angles larger than 90$^\circ$ were rejected, which is
the recommendation of the LAT team to exclude possible contamination
from the Earth's limb.

\subsection{Variability Analysis}\label{sec2.2}

Since pulsar \gr\ emission is stable, compared to the highly variable blazars,
variability analysis can help to select pulsars among blazars for high
Galactic sources. However, FL8Y does not have the variability information.
Therefore, for each target, we performed aperture photometry to construct
its light curve. We followed the steps provided by the \fermi\ Science Support
Center$\footnote{\url{https://fermi.gsfc.nasa.gov/ssc/data/analysis/scitools/aperture\_photometry.html}}$.
An aperture of radius 1$^\circ$ was used.
We obtained light curves binned in 30-day intervals. There were 124 data points
in each light curve, but, as \fermi\ encountered operational anomaly in
March of 2018, one corresponding data point was not included.
Following the LAT catalogs, we estimated the variability index for each target,
{which is defined as the sum of the differences between the log likelihoods,
in which the source flux is constant and those in which the flux is optimized
in each time bin} \cite{1fgl10}.
The criterion was for a $\chi^2$ distribution with 122 degrees of freedom,
a light curve was considered significantly variable if the $\chi^2$ value was
larger than 161.25 (at a 99\% confidence level).
We found that 13 of the 20 new targets did not have significant variations.
Therefore, 13 sources were selected from this analysis.
We also checked the variability indices for the 37~sources
in our previous analysis, and they were all still non-variables.

We note that, in the past few years, it has been found that several
MSP binaries show orbital modulation in $\gamma$-rays, which suggests
that certain flux variations may be seen from MSP binaries. However,
these~MSP binaries either have very weak periodic signals
(e.g., \cite{rom+15,str+16}), or are much brighter than our targets
(e.g., \cite{ng+18,ark18}). In any case, the variability analysis here is mainly
to exclude blazars as completely as possible, since blazars are the dominant
population among the extra-galactic sources.

\subsection{Maximum Likelihood Analysis}

We performed a standard binned maximum likelihood analysis \cite{mat+96}
on the data of each target, using LAT science tools package v11r5p3 with
the P8R2\_SOURCE\_v6 instrument response functions.
A source model, based on the FL8Y catalog, was created, which included
all sources within 20$^\circ$ of a target. The source model also included
the Galactic and the extragalactic diffuse background models,
gll\_iem\_v06.fits and iso\_P8R2\_SOURCE\_V6\_v06.txt \cite{ace+16},
respectively.
The spectral normalization and photon index parameters of
the sources within 4$^\circ$ from
each target were set free, and the normalizations of the
diffuse background components were also free parameters.
All the other parameters were fixed at their catalog values.

We generated $2^\circ\times 2^\circ$-size
TS maps for our targets, for the purpose of checking whether a target was
a point-like source, without being contained in an extended emission region
or mixed with nearby unknown sources. Previously, this step was important
as some sources were found to have such problems and were rejected.
For a given source, the square root of its TS value was,
approximately, the detection significance. We examined the TS maps of our
13 targets, and found that all of them were `clean' point-like
sources without any contamination.
We also repeated the analysis for the 37 previous sources, since more data
were available than before, and they were
confirmed to be clean.

\subsection{Spectral Analysis}
\label{subsec:sa}

To evaluate the curvature significances for our targets, we used the models
of a power law (PL) and a PL with an exponential cutoff (PLE); the latter
is considered to describe pulsar \gr\ emission well, with
the cutoff energies at several GeV \cite{2fpsr13,xw16}.

The PL model has the form
\begin{equation}
\frac{dN}{dE} = N_0 \left(\frac{E}{E_0}\right)^{-\Gamma}\  ,
\end{equation}
where $N_{0}$ is the normalization, $\Gamma$ is the photon index, and we set
$E_{0} = 1$ GeV.
The PLE model has the~form
\begin{equation}
\frac{dN}{dE} = N_0 \left(\frac{E}{E_0}\right)^{-\Gamma}\exp(-\frac{E}{E_c})\ ,
\end{equation}
where $E_{c}$ is the cutoff energy.

Running the \textit{gtlike} task, we obtained
$ L_{PL} $ and $ L_{PLE} $, the maximum likelihood values modeled with PL
and PLE, respectively, for each target. Then, the curvature significance was calculated by
$Signif\_curve=\sqrt{2log(L_{PLE}/L_{PL})}$. We found that, among the 13 new
sources, 12 of them had significant curvature and one, J1722.8$-$0418, did
not; the results, for both cases, are given in Table~\ref{tab:cmsps} and
Table~\ref{tab:nosc}, respectively. It is interesting to note that there
were three sources (in Table~\ref{tab:cmsps}) which were previously analyzed and
rejected, because of either low
curvature significance or not being a clean point-like source.
We also checked our previous 37 candidates
and found that one source, J1544.5$-$1126, did not have significant curvature
anymore. Its spectral results are given in Table~\ref{tab:nosc}.

\begin{figure}[H]
\centering
\includegraphics[width=0.45\textwidth, angle=0]{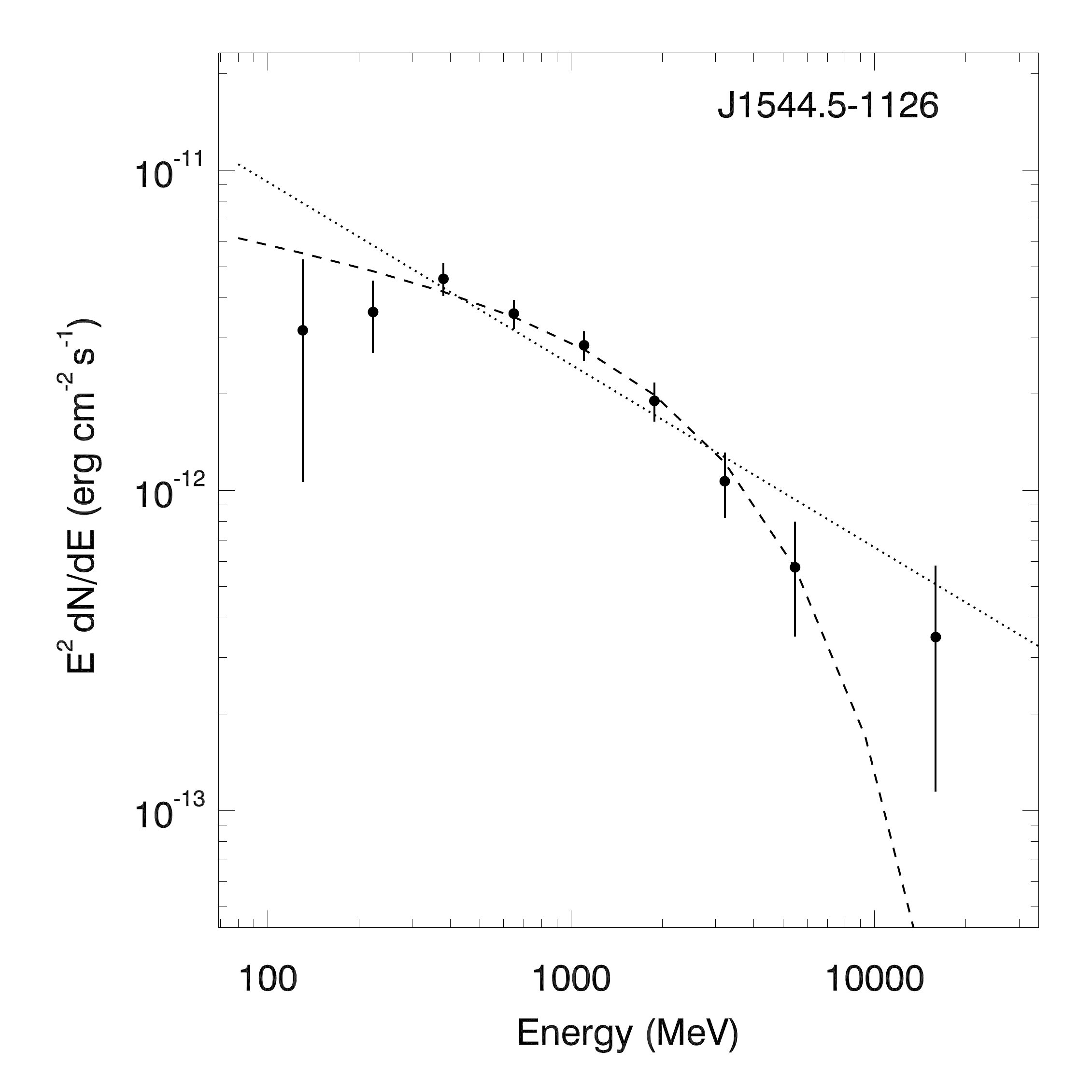}
\includegraphics[width=0.45\textwidth, angle=0]{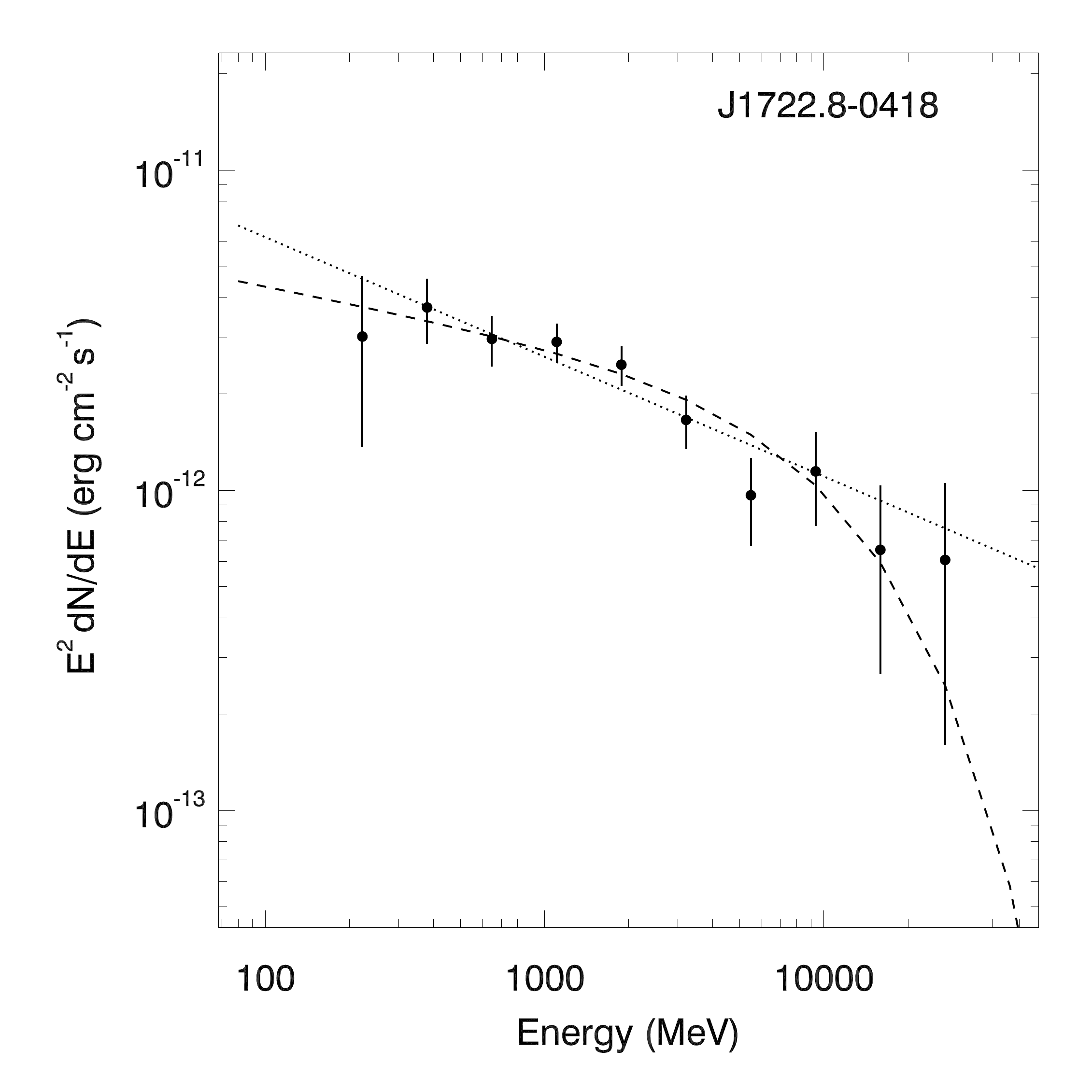}
\caption{The $\gamma$-ray spectra of the sources J1544.5$-$1126 ({\bf left}) and
J1722.8$-$0418 ({\bf right}). The dotted and dashed curves are the best-fit
Power Law (PL) and PL with an exponential cutoff (PLE) models, respectively.}
\label{fig:spec}
\end{figure}

We evenly divided energy logarithmically from 0.1 to 300 GeV into 15 energy
bands, and obtained the spectra of these sources. Spectral data points
with their flux values two times greater than the flux uncertainties were kept.
The spectra of J1722.8$-$0418 and J1544.5$-$1126,
as well as their best PL and PLE fits, are shown in Figure~\ref{fig:spec}.
The spectra confirm our above results.

\begin{table} [H]
		%\bc
\centering
		\begin{minipage}[]{100mm}
	\caption[]{Spectral results for 12 candidate \gr\ millisecond pulsars (MSPs). \label{tab:cmsps}}
		\end{minipage}
	\tiny
	\begin{tabular}{lcccccc}	
		%	\hline
		\toprule
		\textbf{\multirow{2}{*}{FL8Y Name}}  & \textbf{\multirow{2}{*}{Model}}& \textbf{Flux/10\boldmath$^{-9}$}  & \textbf{\multirow{2}{*}{$\Gamma$}} & \textbf{E\boldmath$_{c}$} & \textbf{\multirow{2}{*}{TS}} & \textbf{Signif\_Curve}   \\
		                    &          & \textbf{(photons\,cm\boldmath$^{-2}$s$^{-1}$)} &                   & \textbf{(GeV)}   &             & \textbf{(\boldmath$\sigma$)}      \\
		\midrule
		J0259.0$+$0552              & PL         &       3.7 $\pm$      0.5     &  2.05 $\pm$ 0.06 &                             & 358  &             \\
		                            & PLE &       2.8 $\pm$      0.5     &  1.7  $\pm$ 0.1  &      22    $\pm$      9     & 362  &       3.12  \\
		\midrule
		J0418.9$+$6636              & PL         &       9.5 $\pm$      0.6     & 2.29  $\pm$ 0.04 &                             & 650  &             \\
		                            & PLE &       7.2 $\pm$      0.7     &  1.7  $\pm$ 0.1  &       5    $\pm$      1     & 681  &       6.74  \\
		\midrule
		J0523.3$-$2527 $^\ast$              & PL         &       10.3$\pm$      0.4     & 2.11  $\pm$ 0.03 &                             & 2206 &             \\
		                            & PLE &       7.5 $\pm$      0.5     &  1.3  $\pm$ 0.1  &       3.3  $\pm$      0.5   & 2279 &       11.17 \\
		\midrule
		J0533.7$+$5946              & PL         &       7.3 $\pm$      0.6     & 2.54  $\pm$ 0.06 &                             & 289  &             \\
		                            & PLE &       6.0 $\pm$      0.6     &  1.6  $\pm$ 0.2  &       1.7  $\pm$      0.4   & 326  &       6.45  \\
		\midrule
		J0736.9$-$3231 $^a$ & PL         &       16  $\pm$      1       & 2.62  $\pm$ 0.05 &                             & 505  &             \\
		                            & PLE &      14   $\pm$      1       &  2.1  $\pm$ 0.4  &       2    $\pm$      2     & 509  &       3.53  \\
		\midrule
		J0843.4$+$6713              & PL         &       2.6 $\pm$      0.3     & 1.95  $\pm$ 0.05 &                             & 472  &             \\
		                            & PLE &       1.5 $\pm$      0.2     &  1.0  $\pm$ 0.2  &       3.9  $\pm$      0.7   & 516  &       6.62  \\
		\midrule
		J0940.5$-$7610              & PL         &       7.5 $\pm$      0.6     & 2.36  $\pm$ 0.05 &                             & 392  &             \\
		                            & PLE &       5.8 $\pm$      0.7     &  1.7  $\pm$ 0.2  &       3.1  $\pm$      0.8   & 416  &       5.91  \\
		\midrule
		J1221.5$-$0634              & PL         &       6.2 $\pm$      0.6     & 2.32  $\pm$ 0.06 &                             & 421  &             \\
		                            & PLE &       4.5 $\pm$      0.6     &  1.8  $\pm$ 0.2  &       3    $\pm$      1     & 424  &       4.69  \\
		\midrule
		J1543.6$-$0245 $^a$ & PL         &       7.3 $\pm$      0.6     & 2.51  $\pm$ 0.06 &                             & 266  &             \\
		                            & PLE &       6.3 $\pm$      0.7     &  1.9  $\pm$ 0.2  &       3    $\pm$      1     & 283  &       4.57  \\
		\midrule
		J1603.3$-$6010              & PL         &       3.4 $\pm$      0.4     & 2.01  $\pm$ 0.06 &                             & 190  &             \\
		                            & PLE &       1.4 $\pm$      0.2     &  0.6  $\pm$ 0.1  &       3.2  $\pm$      0.4   & 229  &       6.29  \\
		\midrule
		J1827.5$+$1141  $^b$ & PL         &       6.1 $\pm$      0.8     & 2.31  $\pm$ 0.07 &                             & 238  &             \\
		                            & PLE &       3.7 $\pm$      0.8     &  1.5  $\pm$ 0.2  &       4.4  $\pm$      1.5   & 247  &       4.65  \\
		\midrule
		J2333.1$-$5528              & PL         &       3.7 $\pm$      0.3     & 2.24  $\pm$ 0.06 &                             & 378  &             \\
		                            & PLE &       2.8 $\pm$      0.3     &  1.6  $\pm$ 0.2  &       4    $\pm$      1     & 399  &       5.01  \\

		\bottomrule
	\end{tabular}
	\begin{tabular}{@{}c@{}}
\multicolumn{1}{p{\textwidth -.88in}}{\footnotesize Notes: (1) $a$ denotes the source which was previously excluded for low curvature significance in \citep{dai+16,dai+17}; \mbox{(2) $b$ denotes} the source which was previously excluded for not being a clean point-like source in \citep{dai+16,dai+17}; (3) * denotes that this source is a candidate MSP binary \cite{str+14}, but it is still marked as unidentified in the catalog.}
\end{tabular}

%	\ec
%\begin{minipage}[]{160mm}
%{\small Notes: 1) $a$ denotes the source was previously excluded for low curvature significance in \citep{dai+16}and \citep{dai+17}; 2) $b$ denotes the source was previously excluded for not being a clean point-like source in \citep{dai+16}and \citep{dai+17}; 3) * this source is a candidate MSP binary \cite{str+14}, but it is still marked as unidentified in the catalog.}
%\end{minipage}
\end{table}
\unskip

\begin{table} [H]
\centering
	%	\bc
	%	\begin{minipage}[]{100mm}
	\caption[]{Sources without sufficient curvature significance. \label{tab:nosc}}
	%	\end{minipage}
	\tiny
	\begin{tabular}{ccccccc}	
		%	\hline
		\toprule
		\textbf{\multirow{2}{*}{FL8Y Name}} & \textbf{\multirow{2}{*}{Model}} & \textbf{Flux/10\boldmath$^{-9}$}  & \textbf{\multirow{2}{*}{\boldmath$\Gamma$}} & \textbf{E\boldmath$_{c}$} & \textbf{\multirow{2}{*}{TS}} & \textbf{Signif\_Curve}   \\
		                   &           & \textbf{(photons\,cm\boldmath$^{-2}$s$^{-1}$)} &                   & \textbf{(GeV)}   &             & \textbf{(\boldmath$\sigma$)}      \\
		\midrule	
		J1544.5$-$1126              & PL         &       12.3 $\pm$      0.7     & 2.57  $\pm$ 0.05 &                             & 597.98  &             \\
		                            & PLE &       11.6 $\pm$      0.8     &  2.2  $\pm$ 0.2  &       3    $\pm$      2     & 598.39  &       2.06  \\
		J1722.8$-$0418 & PL         &       11  $\pm$     1     &  2.37  $\pm$ 0.06   &                          & 362 &     \\
		             & PLE &       9   $\pm$     1     &  2.2   $\pm$ 0.1    &       14  $\pm$      7   & 360 &       2.71 \\
		\bottomrule
	\end{tabular}
	%\ec
\end{table}

\section{{{Swift}} X-ray Data Analysis}
We searched for possible X-ray counterparts to the 12 candidates.
Among them, J0523.3$-$2527 has been considered to be an MSP binary system,
and two other sources (J0533.7+5946 and J1543.6$-$0245) have not been covered
by X-ray observations. For the rest of the sources, we obtained the
{Swift} X-ray Telescope (XRT; \cite{bur+05}) data of the
longest exposure from the High Energy Astrophysics Science Archive Research
Center (HEASARC).  The data were processed using the {\sc xrtpipeline} tool in
the HEASOFT package version 6.22.1 distributed by HEASARC, and the calibration
files version 20180710 available in the {Swift} CALDB.
 We searched the X-ray sources in the XRT images using the {\sc detect} command
available in {\sc ximage} with a detection threshold of $3\sigma$.
A detected X-ray source inside the $2\sigma$ {Fermi} error circle was
considered as a possible X-ray counterpart.

Among the 9 sources,
only J0259.0+0552 had a possible X-ray counterpart, while the rest of
the sources had no counterparts. The exact position of the detected X-ray
source was derived using the {\sc xrtcentroid} task, and the source and background
events were extracted from a circular region of radius 47 arcsec.
This source was faint, and there were not sufficient spectral counts
to perform detailed spectral modeling; thus we used the Cash
Statistic \cite{cash79} for the spectral fitting. A simple absorbed power law
was used for the spectral fitting, where the absorption was fixed at
the Galactic value~\cite{kal+05}, and best-fit spectral parameters are given
in Table~\ref{spec}. For the no X-ray counterpart cases, the $3\sigma$ upper
limits on the count rates were estimated, using the {\sc uplimit} command
in {\sc ximage}. We, then, converted the upper limits on the count rates
into fluxes by using webPIMMS\footnote{\url{https://heasarc.gsfc.nasa.gov/cgi-bin/Tools/w3pimms/w3pimms.pl}},
where we assumed an absorbed PL model with $\Gamma_X = 1.7$ and the absorption
column density of the Galactic values. The results are given in
Table~\ref{uplimit}.

\begin{table}[H]
%\begin{center}
%\tabletypesize{\small}
%\tablecolumns{9}
%%\setlength{\tabcolsep}{8.0pt}
%\tablewidth{320pt}
\centering
	\caption{Possible X-ray counterpart to J0259.0+0552.}
	\scalebox{0.9}[0.9]{

 	\begin{tabular}{ccccccccc}
	\toprule

\textbf{\multirow{2}{*}{Source}} & \textbf{\multirow{2}{*}{OBSID}} & \textbf{Exposure} & \textbf{R.A. }& \textbf{Dec.} & \boldmath$N_{\rm H}/10^{20}$ & \multirow{2}{*}{\boldmath$\Gamma_X$} & \boldmath$F_{\rm X}$ & \boldmath$\chi^2/\rm d.o.f$ \\
 & & \textbf{(sec)} & \textbf{(h:m:s)} & \textbf{(\boldmath$^\circ:':"$)} & \textbf{(cm\boldmath$^{-2}$)} & & \textbf{(\boldmath$10^{-13}$)} & \\
\midrule

J0259.0+0552 & 00084646003 & 3094 & 02:58:57.86 & +05:52:44.25 & 8.18 & $2.30^{+0.88}_{-0.66}$ & $4.99^{+1.93}_{-1.62}$ & 2.73/5 \\

\bottomrule

\end{tabular} }
\begin{tabular}{ccc}
\multicolumn{1}{c}{\footnotesize Notes: $F_{\rm X}$ is unabsorbed flux in 0.3--10 keV band in units
of $\rm erg~cm^{-2}~s^{-1}$.}
\end{tabular}

%\begin{minipage}[]{160mm}
%{\small Notes: $F_{\rm X}$ is unabsorbed flux in 0.3--10 keV band in units
%of $\rm erg~cm^{-2}~s^{-1}$.}
%\end{minipage}
\label{spec}
\end{table}
\unskip

\begin{table}[H]
%\begin{center}
%\tabletypesize{\small}
%\tablecolumns{8}
%\setlength{\tabcolsep}{10.0pt}
%\tablewidth{320pt}
\centering
	\caption{X-ray flux upper limits for eight candidate MSPs.}
 	\begin{tabular}{cccccc}
	\toprule

\textbf{\multirow{2}{*}{Source}} & \textbf{\multirow{2}{*}{OBSID}} & \textbf{{Exposure}} & \textbf{Counts} & \boldmath$N_{\rm H}$ & \boldmath$F_{\rm X}^{Upper}$ \\
 & & \textbf{(sec)} & \boldmath$(10^{-3}$ s$^{-1}$) & \textbf{(\boldmath$10^{20}$\,cm$^{-2}$)} & \textbf{(\boldmath$10^{-13}$)} \\
\midrule

J0418.9+6636 & 00047153001 & 3678 & <2.88 & 20.5 & <1.77 \\
J0736.9$-$3231 & 00047171006 & 2394 & <4.68 & 39.2 & <3.52 \\
J0843.4+6713 & 00031603001 & 4617 & <1.89 & 3.28 & <0.84 \\
J0940.5$-$7610 & 00084695002 & 737  & <21.17 & 9.84 & <10.96 \\
J1221.5$-$0634 & 00032457002 & 5071 & <2.93 & 2.42 & <1.27 \\
J1603.3$-$6010 & 00084769008 & 1298 & <9.47 & 24.9 & <6.16 \\
J1827.5+1141 & 00047276001 & 2141 & <4.63 & 14.5 & <2.60 \\
J2333.1$-$5528 & 00084897014 & 1929 & <6.10 & 1.23 & <2.55 \\

\bottomrule

\end{tabular}
\begin{tabular}{ccc}
\multicolumn{1}{c}{\footnotesize Notes: $F_{\rm X}^{Upper}$
is the $3\sigma$ upper flux limit in 0.3--10 keV band in units
of $\rm erg~cm^{-2}~s^{-1}$.}
\end{tabular}

%\begin{minipage}[]{160mm}
%{\small Notes: $F_{\rm X}^{Upper}$
%is $3\sigma$ upper flux limit in 0.3--10 keV band in units
%of $\rm erg~cm^{-2}~s^{-1}$.}
%\end{minipage}
\label{uplimit}
%\end{deluxetable}
%\end{center}
\end{table}

\section{Results and Discussion}

Based on the latest FL8Y catalog and analysis of 10.2 years of \fermi\ LAT
data, we have revisited our selection of candidate MSPs. Now, the number of
catalog sources that fit the MSP type nearly doubles if our simple selection
criteria are used. We have analyzed the data for relatively bright sources
with detection significances greater than 15$\sigma$, and 48 among
57 selected targets have been established as good candidate MSPs. Among
the rejected, the source J1544.5$-$1126 was found not to have sufficient
curvature significance. However, it has been studied at multi-wavelengths
and identified as a $\gamma$-ray-emitting low mass X-ray binary
(a good candidate of the transitional MSP \cite{bhl15,bh15}).
On the other hand, among 12 new candidates, we have recovered
three sources from the rejected ones in our previous analysis. Previously,
we could reject a few sources because they were located in an extended emission
region or mixed with nearby sources, but in this analysis we did not find any such cases.
Combining these facts, the latest catalog appears to be improved.

\begin{figure}[H]
\centering
\includegraphics[width=0.6\textwidth, angle=0]{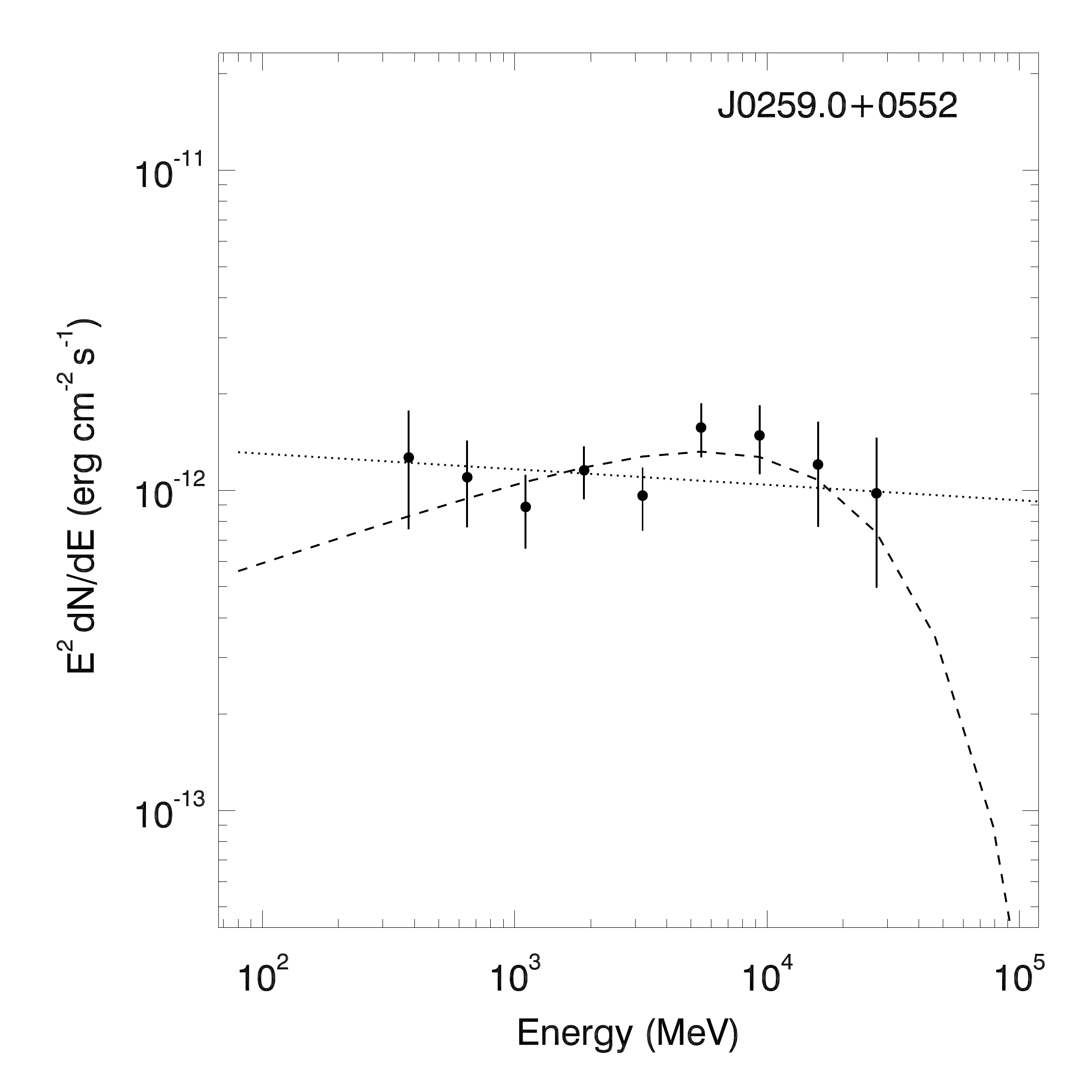}
\caption{The $\gamma$-ray spectrum of the source J0259.0+0552.
The dotted and dashed curves are the best-fit PL and PLE models, respectively.}
\label{fig:spec02}
\end{figure}

Among the new candidates (see Table~\ref{tab:cmsps}), J0523.3$-$2527 is
an actual candidate MSP binary, based on mutli-wavelength studies \cite{str+14}, but
as it is marked as an unidentified source in the catalog, it appears
in our selection. The spectra of pulsars generally have a form of a power
law with an exponential cutoff. The parameter ranges of $\Gamma=1.43$--1.64 and
$E_c=3.00$--4.65 (3$\sigma$) were obtained from the fitting of the \gr\ spectra of
39 known MSPs with a PLE model \cite{xw16}. Considering the ranges,
particularly the $\Gamma$ range which is more reliably determined from
spectral analysis, 10 of the 11 new candidates have consistent values,
suggesting they are good MSP candidates. The exceptional one, J0259.0+0552, has
$E_c \sim 22\pm9$ GeV, which is too large; although the uncertainty is
also large. In~Figure~\ref{fig:spec02}, we show the spectrum of J0259.0+0552, indicating
its strong emission at high energies of $\sim$10 GeV. Given this, it is
not likely a MSP, which has also been pointed out recently \cite{fra+18}.
In addition, we note that J1221.5$-$0634 has been suggested to be associated
with a redshift z = 0.44 quasar \cite{amm+16}, but the association is
questionable~\cite{lp17}. The results are generally consistent with those
obtained from statistical and machine learning techniques \cite{saz+16},
in which most of our new candidates were also classified as MSPs or a young
pulsar (J0736.9$-$3231), but J0259.0+0552, J0843.4+6713, and J1221.5$-$0634
were classified as active galactic nuclei.

The list of the selected candidates are among those being observed at the 13 cm band
(2.2--2.4 GHz)
with the 65-m Tian Ma Radio Telescope (see, e.g., \cite{yan+18}),
which is a fully steerable
Cassegrain antenna.  The~pulsar searching mode, with a time resolution
of 65 $\upmu$s, is being used in our observations. Hopefully the observations,
{a test run with the Tian Ma Telescope,} will
turn out to be fruitful. { If that is the case, we would lower the detection
significances and have more candidates for investigation. The~selection
criteria might also be adjusted to include more candidates.}
In addition, observations at other wavelengths
will certainly help; for example, since the known \gr\ pulsars have  $\gamma$-ray to X-ray flux ratios of 10--10$^5$ \cite{2fpsr13}, X-ray
observations can help to locate source positions accurately. In our analysis of
archival {Swift} X-ray data, only one possible X-ray counterpart to
J0259.0+0552 was found, and for the other 8 sources covered with
the X-ray observations, flux upper limits of
\mbox{$\sim$$10^{-13}$ erg\,s$^{-1}$\,cm$^{-2}$} were obtained. The $\gamma$-ray fluxes
of these sources in 0.1--100 GeV were
(4--10)$\times 10^{-12}$ erg\,s$^{-1}$\,cm$^{-2}$, indicating that, if these
sources are MSPs, the X-ray observations were probably not deep enough.
Thus, further
X-ray observations are also being planned for identifying candidate MSPs
and studying their likely binary nature.

%%%%%%%%%%%%%%%%%%%%%%%%%%%%%%%%%%%%%%%%%%
%\section{Conclusions}

%This section is not mandatory, but can be added to the manuscript if the discussion is unusually long or complex.

%%%%%%%%%%%%%%%%%%%%%%%%%%%%%%%%%%%%%%%%%%
%\section{Patents}
%This section is not mandatory, but may be added if there are patents resulting from the work reported in this manuscript.

%%%%%%%%%%%%%%%%%%%%%%%%%%%%%%%%%%%%%%%%%%
\vspace{6pt}

%%%%%%%%%%%%%%%%%%%%%%%%%%%%%%%%%%%%%%%%%%
%% optional
%\supplementary{The following are available online at \linksupplementary{s1}, Figure S1: title, Table S1: title, Video S1: title.}


%%%%%%%%%%%%%%%%%%%%%%%%%%%%%%%%%%%%%%%%%%
\authorcontributions{formal analysis, D.X.; writing--original draft preparation, D.X.; writing--review and editing, W.Z.; supervision, W.Z.; X-ray analysis, J.V.}

%%%%%%%%%%%%%%%%%%%%%%%%%%%%%%%%%%%%%%%%%%
\funding{This research was funded by the National Program on Key Research
and Development Project (Grant No. 2016YFA0400804) and the National Natural
Science Foundation of China (11633007).}

%%%%%%%%%%%%%%%%%%%%%%%%%%%%%%%%%%%%%%%%%%
\acknowledgments{ This research made use of the High Performance Computing
Resource in the Core Facility
for Advanced Research Computing at Shanghai Astronomical Observatory.}

%%%%%%%%%%%%%%%%%%%%%%%%%%%%%%%%%%%%%%%%%%
\conflictsofinterest{The authors declare no conflict of interest.}

%%%%%%%%%%%%%%%%%%%%%%%%%%%%%%%%%%%%%%%%%%
%% optional
%\abbreviations{The following abbreviations are used in this manuscript:\\

%\noindent
%\begin{tabular}{@{}ll}
%MDPI & Multidisciplinary Digital Publishing Institute\\
%DOAJ & Directory of open access journals\\
%TLA & Three letter acronym\\
%LD & linear dichroism
%\end{tabular}}

%%%%%%%%%%%%%%%%%%%%%%%%%%%%%%%%%%%%%%%%%%
%% optional
%\appendixtitles{no} %Leave argument "no" if all appendix headings stay EMPTY (then no dot is printed after "Appendix A"). If the appendix sections contain a heading then change the argument to "yes".
%\appendix
%\section{}
%\unskip
%\subsection{}
%The appendix is an optional section that can contain details and data supplemental to the main text. For example, explanations of experimental details that would disrupt the flow of the main text, but nonetheless remain crucial to understanding and reproducing the research shown; figures of replicates for experiments of which representative data is shown in the main text can be added here if brief, or as Supplementary data. Mathematical proofs of results not central to the paper can be added as an appendix.

%\section{}
%All appendix sections must be cited in the main text. In the appendixes, Figures, Tables, etc. should be labeled starting with `A', e.g., Figure A1, Figure A2, etc.


%=====================================
% References, variant A: internal bibliography
%=====================================
\reftitle{References}
%%%%\bibliography{psr_ref}
%%%\begin{thebibliography}{999}

%=====================================
% References, variant B: external bibliography
%=====================================
%\externalbibliography{yes}
%\bibliography{your_external_BibTeX_file}

%%%%%%%%%%%%%%%%%%%%%%%%%%%%%%%%%%%%%%%%%%
%% optional
%%\sampleavailability{Samples of the compounds ...... are available from the authors.}

%% for journal Sci
%\reviewreports{\\
%Reviewer 1 comments and authors\E2\80?response\\
%Reviewer 2 comments and authors\E2\80?response\\
%Reviewer 3 comments and authors\E2\80?response
%}

%%%%%%%%%%%%%%%%%%%%%%%%%%%%%%%%%%%%%%%%%%
\end{document}